\documentclass[3p,times,procedia]{elsarticle}
\usepackage{nupha_ecrc}

\volume{00}

\firstpage{1}

\journalname{Nuclear Physics A}

\runauth{}

\jid{nupha}

\jnltitlelogo{Nuclear Physics A}

\usepackage{amssymb}
\usepackage[figuresright]{rotating}
\usepackage{nicefrac}

\newcommand{\beq}{\begin{eqnarray}}
\newcommand{\eeq}{\end{eqnarray}}

\newcommand{\bel}[1]{\begin{eqnarray}\label{#1}}
\newcommand{\eel}{\end{eqnarray}}

\newcommand{\rf}[1]{Eq.~(\ref{#1})}

\newcommand{\rfn}[1]{~(\ref{#1})}

\newcommand{\nn}{\nonumber}

\newcommand{\p}{\partial}

\newcommand{\tr}{\rm tr}

\newcommand{\f}[2]{\frac{#1}{#2}}
\newcommand{\onehalf}{{\nicefrac{1}{2}}}
\newcommand{\ed}{{\varepsilon}}       % energy density
\def\TmnU{T^{\mu \nu}}

\def\n0{n_{(0)}}
\def\e0{\varepsilon_{(0)}}
\def\P0{P_{(0)}}
\def\s0{s_{(0)}}

\def\fplusrsxp{f^+_{rs}(x,p)}

\def\bmu{\beta_\mu}

\def\umU{u^\mu}

\def\pmu{p^\mu}

\def\omnL{\omega_{\mu\nu}}
\def\omnU{\omega^{\mu\nu}}
\def\omnLbar{{\bar \omega}_{\mu\nu}}
\def\omnUbar{{\bar \omega}^{\mu\nu}}

\def\omnUD{\tilde {\omega}^{\mu\nu}}

\def\SmunuU{{\hat \Sigma}^{\mu\nu}}

\def\ubarrp{{\bar u}_r(p)}

\def\usp{u_s(p)}

\def\vbarsp{{\bar v}_s(p)}

\def\vrp{v_r(p)}

\def\g5{\gamma_5}

\def\fminusrsxp{f^-_{rs}(x,p)}

\def\slmnU{S^{\lambda, \mu \nu}}

\begin{document}

\begin{frontmatter}



%% Instructions from Editor: Please use the following \dochead only in the preprint version (e-print arXiv etc.); 
%% use empty \dochead{} when submitting to Nuclear Physics A!
\dochead{XXVIIth International Conference on Ultrarelativistic Nucleus-Nucleus Collisions\\ (Quark Matter 2018)}
%% e.g. \dochead{17th International Conference on Dynamical Processes in Excited States of Solids}

\title{Relativistic hydrodynamics with spin}


\author[a,b]{W.~Florkowski}
\author[c]{B.~Friman}
\author[d]{A.~Jaiswal}
\author[a]{R.~Ryblewski}
\author[c,e]{E.~Speranza}

\address[a]{Institute of Nuclear Physics Polish Academy of Sciences, PL-31342 Krakow, Poland}
\address[b]{Jan Kochanowski University, PL-25406 Kielce, Poland}
\address[c]{GSI Helmholtzzentrum f\"ur Schwerionenforschung, D-64291 Darmstadt, Germany}
\address[d]{School of Physical Sciences, National Institute of Science Education and Research, HBNI, Jatni-752050, India}
\address[e]{Institute for Theoretical Physics, Goethe University, D-60438 Frankfurt am Main, Germany}

\begin{abstract}
A newly proposed framework of perfect-fluid relativistic hydrodynamics for particles with spin $\onehalf$ is briefly reviewed. The hydrodynamic equations follow entirely from the conservation laws for energy, momentum, and angular momentum.  The incorporation of the angular-momentum conservation requires that the spin polarization tensor $\omega_{\mu\nu}$ is introduced. It plays a role of a Lagrange multiplier conjugated to the spin tensor $S^{\lambda, \, \mu\nu}$. The space-time evolution of the spin polarization tensor depends on the specific form chosen for the spin tensor. 
\end{abstract}


\end{frontmatter}

\section{Introduction}
\label{sec:intro}

We report herein on our recent work~\cite{Florkowski:2017ruc,Florkowski:2017dyn}, where a new hydrodynamic framework for particles with spin~$\onehalf$ has been introduced. The main motivation for our studies are first positive measurements of the $\Lambda$--hyperon spin polarization~\cite{STAR:2017ckg,Adam:2018ivw,QMstar}. They have already inspired many theoretical studies analyzing the spin polarization and vorticity formation in heavy-ion collisions~\cite{Karpenko:2016jyx,Xie:2017upb,Li:2017slc,Sun:2017xhx,Lan:2017nye,Xia:2018tes}.  Since relativistic hydrodynamics forms now the main tool to model the space-time evolution of matter created in heavy-ion collisions, it is tempting to include the dynamics of spin polarization into such a framework. This opens a new dimension in the hydrodynamic studies which undergo a very rapid development in recent years~\cite{Florkowski:2017olj,Romatschke:2017ejr}.

%%%%%%%%%%%%%%%%%%%%%%%%
\section{Local equilibrium distribution functions}

Our starting point are local equilibrium distribution functions for particles (plus signs) and antiparticles (minus signs) introduced  in \cite{Becattini:2013fla},
\bel{fplusrsxp}
\fplusrsxp = \f{1}{2m} \ubarrp X^+ \usp,  \quad
\fminusrsxp = - \f{1}{2m} \vbarsp X^- \vrp.
\eel
Here $r,s = 1,2$ are spin indices, $u_r$ and $v_s$ are Dirac bispinors, and $X^{\pm}$ are the four-by-four matrices in the spinor space,
\bel{XpmM}
X^{\pm} =  \exp\left[\pm \xi(x) - \bmu(x) \pmu \right] M^\pm, 
\eel
where
\bel{Mpm}
M^\pm = \exp\left[ \pm \f{1}{2} \omnL(x)  \SmunuU \right].
\eel
Here $\beta^\mu= \umU/T$ and $\xi = \mu/T$,  with the temperature $T$, chemical potential $\mu$, and the fluid four-velocity $\umU$ ($u^\mu u_\mu~=~1$). The quantity $\omnL$ is the spin polarization tensor, while  $\SmunuU$  is the Dirac spin operator, which can be expressed in terms of the gamma matrices, $\SmunuU  = (i/4) [\gamma^\mu,\gamma^\nu]$.  Assuming that $ \omnL \omnU > 0$ and $\omnL \omnUD = 0$, where $ \omnUD$ is the dual spin polarization tensor, we find a simple expression for $M^\pm$~\cite{Florkowski:2017ruc},~\footnote{The case with $ \omnL \omnU < 0$ has been recently discussed in \cite{Florkowski:2018myy}, see also \cite{Prokhorov:2018qhq}.}
\bel{Mpmexp}
M^\pm = \cosh(\zeta) \pm  \f{\sinh(\zeta)}{2\zeta}  \, \omnL \SmunuU  , \qquad \zeta= \frac{1}{2} \sqrt{\f{1}{2} \omnL \omnU} .
\eel
The quantity $\zeta$ can be interpreted as the ratio of the spin chemical potential $\Omega$ and the temperature $T$, namely, $\zeta = \Omega/T$. We note that since $M^\pm$ is an even function of $\zeta$, we can, without loss of generality, choose the positive root  in the last equation above. Then, the direction of the polarization is determined by elements of the polarization tensor $\omega^{\mu\nu}$, for details see~\cite{Florkowski:2017dyn}.

The use of the equilibrium distribution functions\rfn{fplusrsxp} leads to the formula for pressure that becomes in this case a function of $T, \,\mu$ and $\Omega$, i.e., $P=P(T,\mu,\Omega)$. The other thermodynamic functions are obtained through the thermodynamic relations:
\bel{dP}
s = \left.{\f{\p P}{\p T}}\right\vert_{\mu,\Omega}, \quad 
n = \left.{\f{\p P}{\p \mu}}\right\vert_{T,\Omega}, \quad 
w = \left.{\f{\p P}{\p \Omega}}\right\vert_{T,\mu},
\eel
where $s, \,n$ and $w$ are  the entropy, charge, and spin densities, respectively.

\section{Hydrodynamic background equations}

The standard expressions for the charge current, energy-momentum tensor, and the entropy current allow us to calculate these quantities directly from the equilibrium distributions~\cite{Florkowski:2017ruc,deGroot:1980}. The energy-momentum tensor has a perfect-fluid, symmetric form and should be conserved, $\partial_\mu T^{\mu\nu} = 0$. This equation can be split into two parts, one which is longitudinal to the flow vector  $u^\mu$ and the other one which is transverse. In this way we obtain two equations:
\begin{eqnarray}
\partial_\mu [(\ed + P) u^\mu ] &=& u^\mu \partial_\mu P,  \label{scon}  \\
 (\ed + P ) u^\nu \p_\nu  \, u^\mu&=& (g^{\mu \alpha} - u^\mu u^\alpha ) \partial_\alpha P, \label{euler}
\end{eqnarray}
where $\ed$ is the energy density. Evaluating the derivative on the left-hand side of the first equation we find
\bel{snwcon}
T \,\partial_\mu (s u^\mu) + \mu \,\partial_\mu (n u^\mu) + \Omega \,\p_\mu (w u^\mu) = 0. \nn
\eel
The middle term vanishes due to charge conservation, $\p_\mu (n u^\mu)=0$. Hence, in order to have entropy conserved in our system (for the perfect-fluid description we are aiming at), we demand that $\p_\mu (w u^\mu) = 0$. Consequently,  we self-consistently arrive at the conservation of entropy, $\partial_\mu (s u^\mu)~=~0$.

Equations \rfn{scon} and \rfn{euler},  supplemented by the conservation of the spin density $w$,  form a closed system of six equations for six unknowns: $T(x)$, $\mu(x)$, $\Omega(x)$ and three components of $u^\mu(x)$. Since they do not determine the time evolution of the individual components of the spin polarization tensor they can be dubbed the equations for hydrodynamic background.

%%%%%%%%%%%%%%%%%%%%% FIGURE %%%%%%%%%%%%%%%%%%%%%
\begin{figure}[t]
\begin{center}
\includegraphics[width=1.0\textwidth]{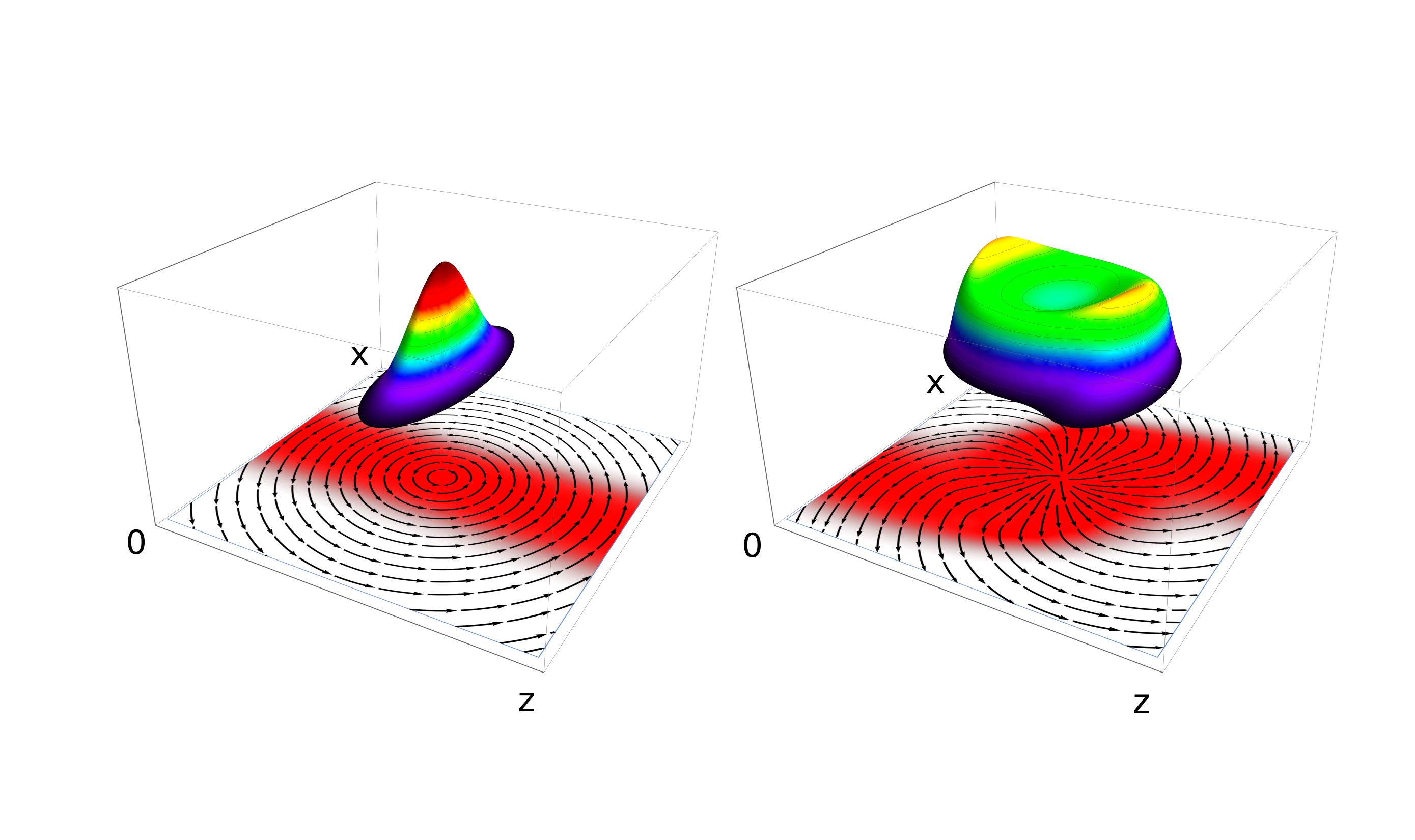}
\end{center}
%\vspace{-0.5cm}
\caption{(Color online) Schematic plots illustrating the space-time evolution of a spin-polarized fluid. Left (right) panel shows the initial (final) system's configuration. The arrows show directions of the hydrodynamic, vortical-like  flow in the reaction plane. The 3D plots in the upper parts show temperature profiles --- the initial profile is elongated along the impact-vector axis, while the final one is rotated and more symmetric due to the outward expansion. The red color in the lower parts marks the region with a given spin polarization which is transported  along the fluid stream lines, according to \rf{st12}. }
%\vspace{-0.3cm}
\label{fig:v}
\end{figure}

%%%%%%%%%%%%%%%%%%%%%%%%%%%%%%%%%%%%%%%
\section{Spin dynamics}

Our approach is based on the conservation of the angular momentum
in the form $\p_\lambda J^{\lambda, \mu\nu}=0$, where $J^{\lambda, \mu\nu} = L^{\lambda, \mu\nu} 
+ S^{\lambda, \mu\nu}$ with $L^{\lambda, \mu\nu}=x^\mu T^{\nu\lambda}- x^\nu T^{\mu\lambda}$
being the orbital angular momentum and $S^{\lambda, \mu\nu}$ being the spin tensor. 
Since our energy-momentum tensor $\TmnU$ turns out to be symmetric, 
the conservation law $\p_\lambda J^{\lambda, \mu\nu}=0$ implies a separate conservation
of the spin tensor $\slmnU$~\cite{Hehl},
\bel{spincon1}
\p_\lambda \slmnU = 0.
\eel
For $\slmnU$ we use the form discussed in \cite{Becattini:2010}
\bel{st11}
\slmnU = \!\!\int\!\!\f{d^3p}{2 (2\pi)^3 E_p} \, p^\lambda \, {\tr} \left[(X^+\!-\!X^-) \SmunuU \right]  =  \frac{w u^\lambda}{4 \zeta}  \omega^{\mu \nu} .
\eel

Using the conservation law for the spin density and introducing the rescaled 
spin polarization tensor $\omnUbar = \omnU/(2\zeta)$, we find
\bel{st12}
u^\lambda \p_\lambda \, \omnUbar = \f{d\omnUbar }{d\tau} = 0.
\eel
Since $\omnUbar$ is antisymmetric, \rf{st12} with the normalization condition 
\bel{norm}
\omnLbar \, \omnUbar = 2
\eel
yields five independent equations. One may also check that if the condition $\omnL \omnUD = 0$ is fulfilled on the initial hypersurface, it is fulfilled at later times, provided \rf{st12}  holds. Hence, \rf{st12}  delivers four extra equations that  are necessary to determine the space-time evolution of a spinning fluid.  We note that \rf{st12} excludes mixing of various components of the spin polarization tensor. A heuristic derivation of~Eqs. \rfn{spincon1} and \rfn{st11} has been recently given within the kinetic-theory approach in \cite{Florkowski:2018ahw}.

In Ref.~\cite{Florkowski:2017ruc} we have shown that our framework has  a vortex solution that corresponds to global equilibrium studied in~\cite{Becattini:2013fla}. In this case, the spin polarization tensor  $\omega$ is equal to the thermal vorticity $\varpi$ (the latter being a rotation of the $\beta$ field),
\bel{tv}
\omega_{\mu\nu} = \varpi_{\mu\nu} = -\f{1}{2} \left(\p_\nu \beta_\mu - \p_\mu \beta_\mu \right).
\eel
In Fig.~\ref{fig:v} we show schematic plots of our preliminary numerical  simulations~\cite{FFJRS} that describe a vortex-like behaviour of the polarized fluid. In this case, \rf{tv} does not hold in general. One expects that the relation~\rfn{tv} is a consequence of a dissipative spin-orbit interaction, which is so far missing in our framework~\cite{Montenegro:2017lvf}. The spin-orbit interaction should lead also to asymmetric energy-momentum tensor~\cite{Yang:2018lew}.

\section{Summary}
\label{sec:sum}

A new hydrodynamic approach to relativistic perfect-fluid hydrodynamics of particles with spin $\onehalf$ has been introduced. The system of hydrodynamic equations follows directly from the conservation laws for electric charge, energy, momentum and angular momentum.  An important ingredient of our approach is the form of the spin tensor defined by \rf{st11} that allows for construction of a consistent system of equations. We note that the form \rfn{st11} is different from those used in \cite{Becattini:2013fla} and \cite{deGroot:1980}, which, by the way,  differ from each other. The role played by the definition of the spin tensor for the formulation of hydrodynamics will be the subject of a separate publication~\cite{BFS}.     

\medskip
{\bf Acknowledgments} This research was supported in part by the ExtreMe Matter Institute EMMI at the GSI 
Helmholtzzentrum f\"ur Schwerionenforschung,  Darmstadt, Germany and by the Polish National Science Center Grant  No. 2016/23/B/ST2/00717. A.~J. was supported in part by the DST-INSPIRE faculty award. E.~S.  was  supported  by  BMBF  Verbundprojekt 05P2015 - Alice at High Rate and by the  Deutsche  Forschungsgemeinschaft  (DFG)  through  the  grant  CRC-TR 211 ``Strong-interaction matter under extreme conditions''.



%% References


% \bibliographystyle{elsarticle-num}
% \bibliography{<your-bib-database>}


% \vspace{-0.2cm}

\end{document}